\documentstyle[twoside,fleqn,espcrc2,psfig]{article}

\newcommand{\AmS}{{\protect\the\textfont2
  A\kern-.1667em\lower.5ex\hbox{M}\kern-.125emS}}

\title{High Energy Gamma Rays }

\author{R. Mukherjee\address{Department of Physics \& Astronomy, \\
        Barnard College, Columbia University, New York, NY 10027}}
       
\begin{document}

\begin{abstract}
This article reviews the present status of high energy $\gamma$-ray 
astronomy 
at energies above 30 MeV. Observations in the past decade using both space- 
and ground-based experiments have been primarily responsible for giving 
a tremendous boost to our knowledge of the high energy Universe. High 
energy $\gamma$-rays have been detected from a wide range of Galactic and 
extragalactic astrophysical sources, such as $\gamma$-ray bursters, pulsars, 
and active galaxies. These observations have established 
high energy $\gamma$-ray astronomy as a vital and exciting field, 
that has a bright future. This review summarizes the experimental techniques, 
observations and results obtained with recent experiments, and concludes with 
a short description of future prospects. 

\vspace{1pc}
\end{abstract}

\maketitle

\section{Introduction}

High energy $\gamma$-rays are one of the most direct ways of studying the 
non-thermal Universe. Sources of very high energy radiation enable us 
to explore the highest energy accelerators in the cosmos, in situations 
with extreme gravitational and magnetic fields. Although high energy 
$\gamma$-rays have been studied for some time, rapid development in the field 
came only in the 1990s. 

The most significant thrust to high energy astronomy above 30 MeV came with the 
launch of the {\sl Compton Gamma Ray Observatory} (CGRO) in 1991, which 
carried onboard the Energetic Gamma-Ray Experiment Telescope (EGRET), along 
with three other instruments. 
Until June 2000, when CGRO was de-orbitted, EGRET produced a wealth 
of astrophysical results, and was responsible for the detection of more than 
270 point sources of $\gamma$-rays \cite{hart99}. 
The space-based results were complemented by 
ground-based atmospheric Cherenkov telescopes (ACTs), which detected 
$\gamma$-rays with energies above 250 GeV from a few of the same sources. 
Particularly, the detection of the Crab Nebula above 250 GeV 
by more than eight separate experiments has firmly established ground-based 
$\gamma$-ray astronomy on a solid foundation \cite{cata99}. 

It is impossible to do justice to the exciting developments in high energy 
$\gamma$-ray astronomy in the few allotted pages. 
This article, therefore, selectively reviews a few highlights of the field, 
and refers the reader for more details to some 
of the more comprehensive reviews that have appeared in the literature 
recently \cite{cata99,ongr98,ongr00,hoff99}. The review is organized as 
follows: 
Sections 2 and 3 of this article summarize the techniques and results from 
EGRET and the ground-based experiments, respectively. While the experimental 
techniques are distinct, there is considerable overlap in the scientific issues
addressed by the space- and ground-based experiments. The three 
other $\gamma$-ray experiments, BATSE, OSSE, and COMPTEL onboard CGRO are not 
included in this review, but are described elsewhere \cite{gehr97}. 
Some recent developments in ground-based experiments, and future 
directions are given in section 4.

\section{ Space-based Experiments}

\subsection{The EGRET Instrument}

EGRET was primarily a spark chamber instrument that detected $\gamma$-rays in 
the energy range $\sim$ 30 MeV to 30 GeV using the pair production technique. 
The instrument had a lifetime of approximately 9 years, from 1991 May to 
2000 June, and proved to be immensely successful. 
It had the standard components of 
a high-energy $\gamma$-ray instrument: an anticoincidence dome to discriminate 
against charged particles, a spark chamber particle track detector with 
interspersed high-$Z$ material to convert the $\gamma$-rays into 
electron-positron pairs, a triggering telescope to detect the presence 
of the pair with the correct direction of motion, and an energy measurement 
system, which in the case of EGRET was constructed of NaI(Tl) crystals. 
The anticoincidence 
dome was successful in rejecting the charged particle background, which 
outnumber the $\gamma$-rays by a factor of $10^4$. EGRET had an 
effective area of 1500 cm$^2$ in the energy range 0.2 GeV to 1 GeV, decreasing 
to about one-half the on-axis value at $18^\circ$ off-axis. 
The threshold sensitivity of EGRET ($> 100$ MeV) for a single 2-week 
observation was 
$\sim 3\times 10^{-7}$ photons cm$^{-2}$ s$^{-1}$. Details of the EGRET 
instrument are given elsewhere \cite{thom93}. 

Prior to EGRET, two other successful satellite experiments, SAS-2 and COS-B, 
pioneered the field of high energy $\gamma$-rays. These experiments were
responsible for making some of the first maps of the $\gamma$-ray sky and
producing the first $\gamma$-ray source catalogs \cite{swan81}. 
 
\subsection{Results from EGRET}

\subsubsection{Point Sources}

The $\gamma$-ray sky in the 30 MeV to 30 GeV range comprises both 
diffuse radiation as well as point sources. The diffuse radiation, as described
in the following section, has 
a Galactic as well as an extragalactic component, 
and needs to be modeled in order to 
do point source analysis of the EGRET data. The point sources detected 
by EGRET above 100 MeV are shown in Fig. 1. 
EGRET has seen several different kinds of sources, as shown in Table 1. In 
addition to the sources listed in the table, EGRET has also detected 5 
$\gamma$-ray bursts \cite{ding98}, the X-ray binary Cen X-3 \cite{vest97}, 
and 1 solar flare \cite{kanb93}. 

\begin{figure}
\psfig{file=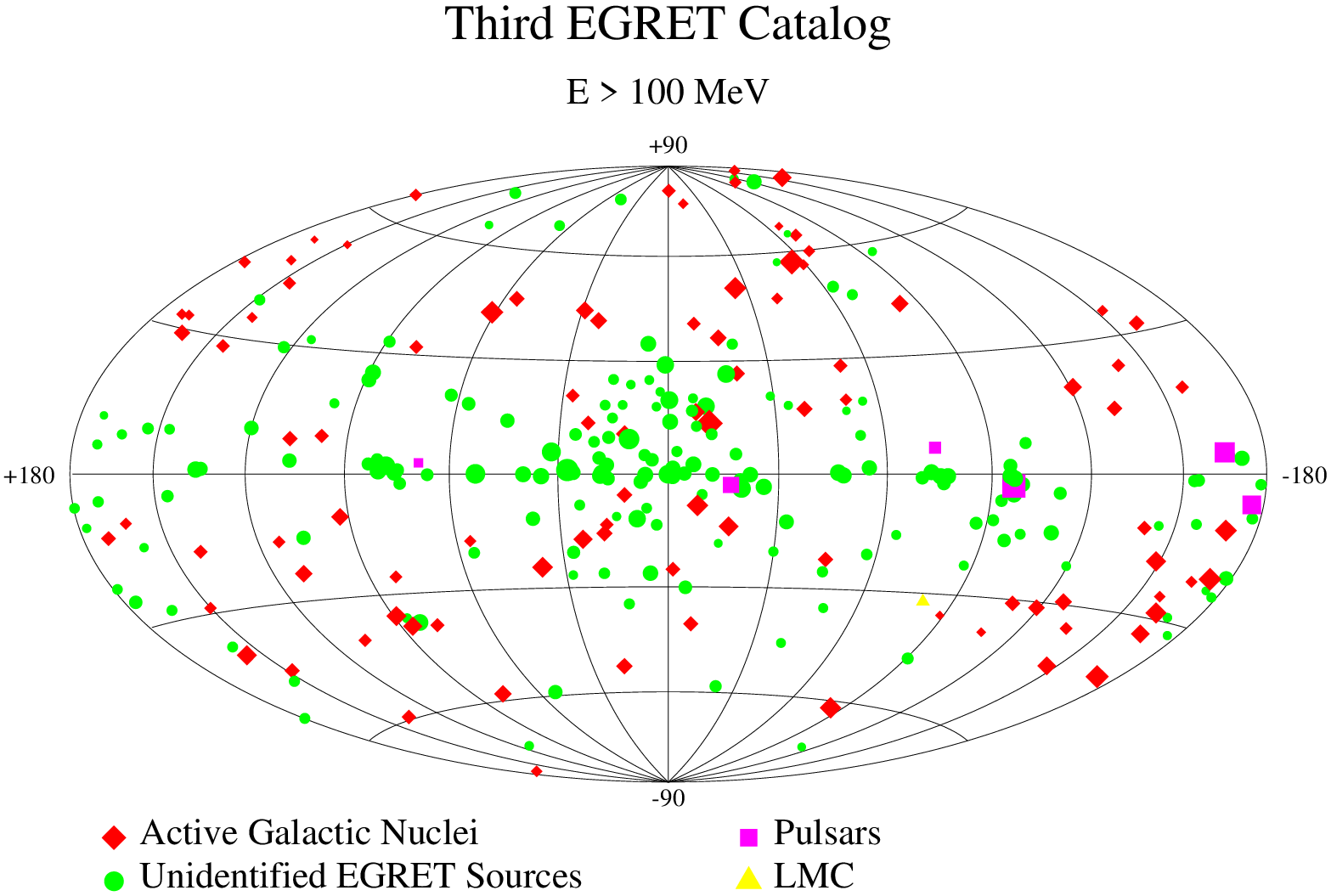,height=2.0in,width=3.0in,bbllx=100pt,bblly=500pt,bburx=750pt,bbury=870pt,clip=.}
\caption{Point sources detected by EGRET at $> 100$ MeV (Third EGRET Catalog) 
\cite{hart99}}
\end{figure}

It is interesting to note that more than 60\% of the EGRET sources are 
unidentified, with no firmly established counterparts at other wavelengths. 
Some of these sources have remained unidentified since the first surveys of 
the $\gamma$-ray sky with the COS-B satellite \cite{swan81}, and are one of 
the outstanding mysteries of the EGRET mission. Some researchers have found 
correlations between unidentified EGRET sources in the Galactic plane and 
supernova remnants \cite{yadi97,stur95,espo96}, while others report 
correlations with OB associations and massive stars \cite{kaar96,rome99}. 
A review of the 
various efforts to understand the nature of these sources is given in 
\cite{mukh97a}. Recently, it has been noted that the mid-latitude unidentified 
EGRET sources form a population distinctly different from the sources along 
the Galactic plane \cite{gehr00,gren00}. Further, it has been suggested that 
these mid-latitude sources 
are probably associated with the Gould Belt of massive stars and gas clouds, 
about 600 light years away \cite{gehr00}. Counterparts to 
individual unidentified sources may only be found after the launch of future,  
more sensitive $\gamma$-ray missions such as GLAST (see section 4).  

\begin{table}
\small
\caption{Sources in the Third EGRET (3EG) Catalog}
\label{table:1}
\begin{tabular}{@{}ll}
\hline
Source Class  &  Number         \\ 
\hline
Blazars       &  67 (94)$^*$    \\ 
Pulsars       &  6              \\ 
Unidentified  &  170            \\ 
Normal Galaxy &  1 (LMC)        \\ 
Radio Galaxy  &  1 (Cen A)      \\ 
\hline
\end{tabular}\\[2pt]
* 27 AGN have been marginally identified.
\end{table}

The majority of the sources away from the
Galactic plane are extragalactic and have been identified with active galactic
nuclei (AGN). Other than the unidentified sources, AGN constitute 
the largest class of EGRET sources. EGRET tends to detect the ``blazar'' class 
of AGN, which are characterized by emissions that include high 
radio and optical polarization, and rapid flux variability at all 
wavelengths.  A large fraction of these sources exhibit apparent superluminal 
motion, as evidenced from VLBI radio observations. 
EGRET has observed variability in several blazars on timescales of days to 
months. A review on the blazars observed by EGRET may be found in 
\cite{hart97}. 

The $\gamma$-ray spectrum of blazars in the 30 MeV to 30 GeV range is
well-described by a single power-law of spectral index $\sim 2.0$. 
The broad-band spectral energy distribution (SED) of one 
EGRET flat-spectrum radio quasar (FSRQ), PKS 0528+134, is shown in Fig. 2. 
The figure demonstrates the typical characteristics of the SED of a quasar 
observed by EGRET, 
showing that the $\gamma$-ray luminosity dominates the spectrum. In fact, in 
most FSRQs the $\gamma$-ray power dominates or equals the power at 
other wavebands. The apparent luminosity, assuming isotropic emission, is 
$10^{48} - 10^{50}$ ergs s$^{-1}$ \cite{mukh97b}. 
Fig. 2 includes broadband observations 
at several different epochs and demonstrates another important characteristic
of EGRET blazars -- spectral variability between different states of the 
source. 

\begin{figure}[h!] 
\psfig{file=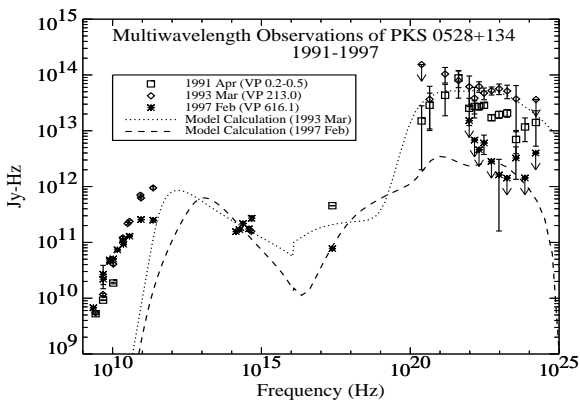,height=2.1in,width=3.0in,bbllx=50pt,bblly=105pt,bburx=560pt,bbury=510pt,clip=.}
\caption{SEDs of the blazar PKS 0528+134 during several different epochs \cite{mukh99}}
\end{figure}

EGRET observations of blazars help to constrain models of emission for these
objects. Rapid variability, high compactness, and superluminal motion suggest 
that the emission is likely to be beamed and Doppler-boosted into the line 
of sight \cite{schl96}. Simplifying many details, it is generally believed that AGN are 
powered by accretion of matter on to a supermassive black hole. 
For the case of blazars, collimated jets composed of shocks or 
plasma blobs move with relativistic speeds outward along the axes of the 
accretion disk, making a small angle with the axis. 

Gamma-ray emission models for blazars are generally divided into two broad 
classes: leptonic and hadronic. Details of these models are not discussed here
but may be found in \cite{hart97}. Fig. 2 shows an example of a leptonic 
jet model applied to the case of PKS 0528+134 \cite{mukh99}. 
It has been suggested that hadronic 
models may be distinguished from leptonic models by the observation of high 
energy neutrinos that may be produced as a result of photoproduction
\cite{muec00}. There are several outstanding questions regarding the 
details of the different models that may only be resolved in the future. 

In addition to blazars, EGRET has detected $\gamma$-rays from 6 pulsars. These 
are the Crab, Vela, PSR 1706-44, B1951+31, B1055-58, and Geminga. In addition, 
soft $\gamma$-rays have been seen from the pulsar B1509-58, although no pulse 
was detected in the EGRET energy range. A few other pulsars have been 
marginally identified. All strongly-detected 
$\gamma$-ray pulsars are double-pulsed, although 
the light curves exhibit a wide variety of patterns. Unlike blazars, the flux 
history of pulsars show relatively constant emission over time. The 
$\gamma$-ray spectra of these sources have power-law indices in the range 
-1.39 to -2.07. Geminga is
an interesting pulsar, as it is radio-quiet, and its period was originally
derived from X-ray measurements. An up-to-date review on EGRET 
pulsars and how they impact models of pulsar $\gamma$-ray emission may be 
found in \cite{thom00}. 

One of the most exciting results in high energy astrophysics is undoubtedly 
the detection of $\gamma$-ray bursts (GRB). More than 2500 bursts have been
recorded by BATSE, but the detection of 5 of these bursts by EGRET 
has demonstrated that GRB are not necessarily a low-energy $\gamma$-ray
phenomena. A review on the high energy emission from GRB is presented elsewhere
in these proceedings \cite{waxm00}. EGRET-detected bursts are described in 
\cite{ding98}. 

\subsubsection{Diffuse Radiation}

EGRET observations of the $\gamma$-ray sky above 30 MeV have provided a 
unique 
view of the diffuse $\gamma$-ray radiation, which constitutes the bulk of the 
emission detected by EGRET. The diffuse emission is found to have a 
Galactic component arising from cosmic-ray interactions 
with the local interstellar gas and radiation \cite{hunt97}, as well as an 
almost uniformly distributed component believed to originate outside the 
Galaxy \cite{sree98}. The average spectrum of the extragalactic diffuse 
emission is well-described by a single power-law 
with an index  $-(2.10\pm 0.03)$ in the 30 MeV to 100 GeV range. 

The precise origin of the extragalactic diffuse emission is not well-known, and
possibly includes both diffuse origin as well as contributions from unresolved 
point sources. These are discussed in \cite{sree98}. Recent theories include 
contributions by 
the upscattering of cosmic microwave background radiation by highly 
relativistic electrons, formed as a result of gravity-induced shock waves 
during the formation of large scale structure in the intergalactic medium
\cite{loeb00}. Models based on discrete source contributions have considered 
an
array of different source classes. Of these the most popular have been 
unresolved blazars. Recent estimates of blazar contribution based on a 
$\gamma$-ray evolution function derived from $\gamma$-ray data is roughly 
25 to 30 percent 
\cite{chia98}, although earlier estimations have found a larger 
contribution \cite{stec96}.  

\section{ Ground-based Experiments}

\subsection{Instruments ($> 300$ GeV)}

At very high energies satellite experiments are made impractical due to the 
rapidly falling photon fluxes. However, at these energies $\gamma$-ray 
astronomy 
can be done using ground-based instruments by detecting the secondary  
particles or radiation produced as a result of the interaction of high energy 
$\gamma$-rays in the upper atmosphere. At energies $> 300 $ GeV, imaging 
atmospheric Cherenkov telescopes (ACTs) have been used successfully to image 
the flashes of Cherenkov light produced as a result of the electromagnetic
cascades initiated by high energy $\gamma$-rays. The field was pioneered by 
the Whipple 10-m telescope, which was the first to detect TeV $\gamma$-rays 
from an astrophysical source \cite{week89}. Imaging ACTs detect the Cherenkov 
flashes using fast photomultipliers (PMTs) located at the focal plane of 
large optical reflectors. The PMT signals are subsequently read out using a 
fast electronics chain. In comparison to satellite experiments, ground-based 
ACTs have much larger collection areas, and better angular resolution, 
although much smaller fields of view. ACTs can only operate on 
cloudless, moonless nights, and thus have a lower duty cycle. 

The Whipple Observatory was followed by several successful high resolution 
ACTs, e.g., CAT, CANGAROO, Durham, GT-48, 
HEGRA, Pachmari, SHALON, TACTIC, Telescope Array, among others. A complete 
list of current telescopes and their properties may be found in 
\cite{ongr98,hoff99}. Air shower arrays, which operate above 30 TeV, and 
directly detect the secondary particles in the air showers are not included in
the review (see \cite{hoff99} for a discussion). 

\subsection{Results}

Figure 3 shows the skymap of point sources detected by
ground-based experiments at energies $> 250$ GeV. In addition to those shown in
the figure, a few other sources have been detected at a lower significance
\cite{week99}. The source list consists of both Galactic, as well
extragalactic objects. 

\begin{figure}
\psfig{file=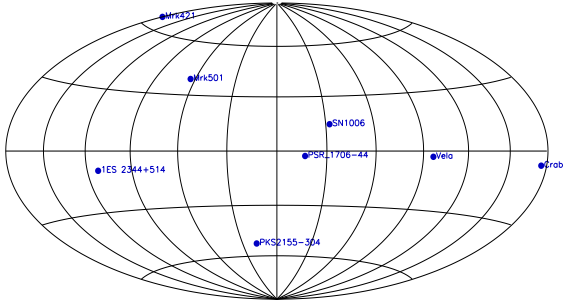,height=1.6in,width=3.0in,bbllx=190pt,bblly=160pt,bburx=815pt,bbury=475pt,clip=.}
\caption{Locations of the $\gamma$-ray point sources in Galactic coordinates 
detected by ground-based ACTs (250 GeV to 15 TeV). }
\end{figure}

The Crab Nebula was the first source to be detected unambiguously in TeV 
$\gamma$-rays. 
Since then it has been detected several times with various different
instruments  and is clearly the standard candle in the 300 GeV to 3 TeV range 
\cite{week99}. No pulsed emission has been observed from the Crab at TeV
energies. In addition to the Crab, two other plerionic supernova remnants 
(SNRs), Vela and PSR 1706-44, and 
a couple of shell-type SNRs have been detected by ACTs 
(see \cite{cata99,week99} for a review). 

Mrk 421 was the first extragalactic source to be detected at $> 250$ GeV
\cite{punc92}, and has been observed by several different ground-based
instruments \cite{cata99}. At a redshift of 0.031, it is the closest BL Lac 
object seen by EGRET.  The source exhibits extremely rapid variability
at TeV energies with doubling times of the order of 1 hour or less
\cite{gaid96}. Its TeV flux can vary from a quiescent level of 0.3 to 
$10\times$ Crab flux during a $\gamma$-ray flare. Mrk 421 exhibits the 
fastest time scale $\gamma$-ray variability 
seen in any blazar to date. Figure 4 shows the spectacular flare from the
source in 1996. Cross-correlations of various data sets of 
Mrk 421 indicate a significant correlation of the X-ray and TeV 
$\gamma$-ray flux variability \cite{buck96}.  

\begin{figure}[h!] 
\psfig{file=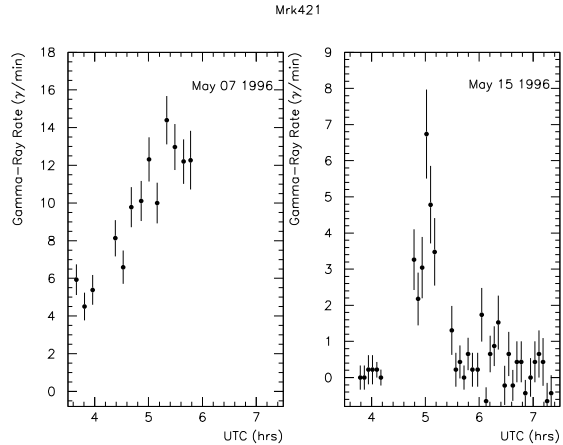,height=2.0in,width=3.0in,bbllx=50pt,bblly=450pt,bburx=545pt,bbury=775pt,clip=.}
\caption{Flux of TeV photons from Mrk 421 as a function of time for two 
separate flares, indicating variability on the time scale $<$ few hours 
\cite{gaid96}}
\end{figure}

At a redshift of 0.033, Mrk 501 is the second closest X-ray-selected
BL Lac (XBL) known and was the second blazar discovered at TeV energies 
by the Whipple group \cite{quin96}. It has since been observed by HEGRA, CAT, 
TACTIC, and the Telescope Array Project \cite{brad97,punc97,bhat97a,haya98}. 
The detection of Mrk 501 in the EGRET data came only after the 
TeV detection \cite{kata99}, making it the first object to be
discovered as a $\gamma$-ray source from the ground.  The broadband spectrum 
of Mrk 501 during the outburst of 1996 May is shown in Fig. 5 \cite{kata99}. 
EGRET reported a GeV outburst from the source when Mrk 501 was detected at 5.3 
$\sigma$ above 500 GeV \cite{sree00}. 
Mrk~501 also shows significant variability at X-ray, low-energy $\gamma$-ray 
and TeV energies. This source was detected for the first time
in the 100 keV to 1 MeV range by OSSE, the resulting SED showing
this emission to be most likely of synchrotron origin \cite{cata97}. 
This represents the highest synchrotron cutoff seen in any blazar at present. 
At TeV energies, the source exhibited an extraordinary outburst in 1997 
\cite{ahar99}, when its intensity was observed to be 
several times that of the Crab. 

\begin{figure}
\psfig{file=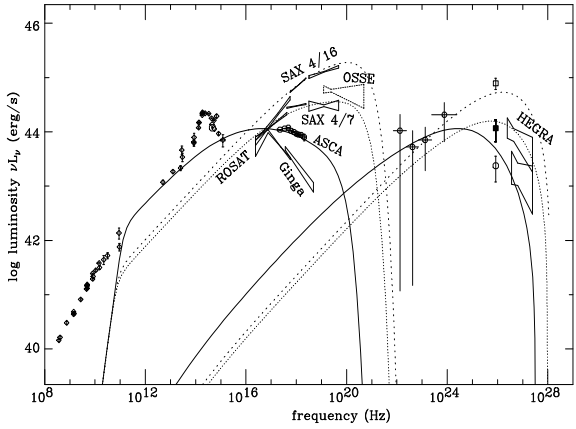,height=2.3in,width=3.0in,bbllx=35pt,bblly=215pt,bburx=550pt,bbury=580pt,clip=.}
\caption{Broadband SED of Mrk 501 during several different epochs. The filled 
and open squares correspond to Whipple flux on 1997 April 7 and 16, 
respectively. The circles correspond to data taken in 1996 March. The diamonds 
correspond to non-simultaneous data from the NED data base. 
\cite{kata99}}
\end{figure}

A comprehensive review of the other sources detected at TeV energies is given 
in \cite{cata99}. 
A comparison of the GeV and TeV skymaps show that only a handful of the EGRET 
sources have been detected at TeV energies, although a simple extrapolation 
of the EGRET spectra predict flux levels well 
above the sensitivity of current atmospheric Cherenkov telescopes. The cutoffs
in the spectra could be either due to intergalactic absorption or intrinsic 
absorption at the source. $\gamma$-rays traversing intergalactic distances 
may be 
absorbed by photon-photon pair production ($\gamma \gamma\to e^+ e^-$), with a
TeV photon preferentially interacting with 0.5 eV (IR) background photon. The
presence of this extragalactic background light (EBL) is probably the main 
explanation for the lack of detection of the majority of EGRET sources at TeV
energies. The spectral cutoffs of the sources is expected to be in the 50 to 
250 GeV range, and is one of the primary motivations for exploring this 
regime. The study of the EBL is important for several reasons. 
Direct measurements of the EBL are sparsely sampled and 
fraught with uncertainties due to the presence of foreground IR radiation 
\cite{haus98,puge96}. The TeV spectra of
Mrk 421 and Mrk 501 have been used to indirectly derive limits on the EBL
\cite{bill98,stan98}. Clearly, more spectral measurements of AGN
at different redshifts, particularly in the 50 to 250 GeV range, 
are required in order to set restrictive limits on the EBL. 

\section{The Future of High Energy Gamma-ray Astronomy}

The future prospects of high energy astronomy are very promising. Recent 
results from space- and ground-based experiments have raised numerous important
questions about the emission mechanisms in astrophysical sources. There is also
the tantalizing possibility of probing novel astrophysical phenomena which 
could arise as a result of new physics beyond the Standard Model 
\cite{jung96}. 
In the future, it is clear that one needs more 
sensitive instruments, both satellite-borne as well as ground-based. It is 
particularly important to explore the energy region between 20 and 250 GeV. 

Recently, two experiments STACEE and CELESTE have demonstrated that lower 
energy thresholds can be achieved by using existing 
large arrays of solar heliostat mirrors to collect Cherenkov light. 
Both these experiments have reported
the detection of the Crab nebula \cite{oser99,naur99} and are in the final
stages of construction. A third experiment, Solar-2  is 
currently being built \cite{tume99}. 
For the first time, the energy gap between space- and 
ground-based experiments is starting to be bridged. 
Experiments that will be built in the future include MAGIC 
in Spain, HESS in Namibia, VERITAS in Arizona, 
and Super-CANGAROO in Australia (see \cite{cata99} and references
therein). The project GRACE in India will involve four independent 
experimental systems that will span nearly ten decades of photon energy 
($\sim 10$ keV - 100 TeV) and do coordinated studies of $\gamma$-ray sources 
\cite{bhat97b}.  These experiments will complement 
each other and together span the energy range from 20 GeV to 10 TeV. 
In addition to the ACTs, MILAGRO, a new kind of $\gamma$-ray detector 
with a large field-of-view and continuous operation has recently
come on line. A prototype
of this detector, Milagrito, reported the possible detection of 
the correlated TeV $\gamma$-ray burst GRB 970417a \cite{atki00}, as well as 
Mrk 501 \cite{atki99}. 

The ground-based experiments will be complemented by two next generation space
experiments, GLAST and AGILE. GLAST, which is expected to be launched in 2005,
is projected to be a state-of-the art detector which will use a Si-strip 
tracker and a CsI calorimeter \cite{gehr99}. GLAST's sensitivity will be a 
factor of $>30$ better than EGRET, and is expected to detect two orders of
magnitude more sources than EGRET. GLAST will have some overlap with
ground-based instruments in the 30-300 GeV regime. 
A smaller EGRET-sized experiment, the Italian AGILE 
\cite{tava99}, sensitive in 30 MeV to 50 GeV energy range, is expected to be 
launched in 2002 and bridge the gap before GLAST. Experiments in high energy 
$\gamma$-rays will be complemented by neutrino and cosmic ray experiments 
\cite{ongr00}, thus promising significant scientific progress in the future. 
\medskip

I wish to thank M. Catanese, J. Kataoka, R. Ong, and P. Sreekumar
for providing some of the data and figures presented here, as well as at the 
Neutrino 2000 conference, and D. Hanna and B. Dingus 
for reading the manuscript. 
This research is supported in part by the National Science Foundation and by an
award from Research Corporation.


\begin{thebibliography}{9}

\bibitem{hart99} R. C. Hartman, et al., ApJ, 123 (1999) 79.  
\bibitem{cata99} M. Catanese \& T. C. weekes, astro-ph 9906501 (1999). 
\bibitem{ongr98} R. A. Ong, Phys. Rep., 305 (1998) 95. 
\bibitem{ongr00} R. A. Ong, astro-ph 0003014 (2000). 
\bibitem{hoff99} C. M. Hoffman, et al., Rev. Mod. Phys., 71 (1999) 897. 
\bibitem{gehr97} N. Gehrels \& C. Shrader, AIP Conf. Proc. 410 (1997) 3. 
\bibitem{thom93} D. J. Thompson, et al., ApJS, 102 (1995) 259. 
\bibitem{swan81} B. N. Swanenburg, et al., ApJ, 243 (1981) L69. 
\bibitem{ding98} B. L. Dingus, J. R. Catelli, E. J. Schneid, AIP
                 Conf. Proc. 428 (1998) 349. 
\bibitem{vest97} W. T. Vestrand, P. Sreekumar, \& M. Mori, ApJ, 483 (1997) 
L49. 
\bibitem{kanb93} G. Kanbach, et al., A\&AS, 97 (1993) 349. 
\bibitem{yadi97} I.-A. Yadigaroglu \& R. W. Romani, ApJ, 476 (1997) 356.  
\bibitem{stur95} S. J. Sturner \& C. D. Dermer, A\&A, 281 (1995) L17.
\bibitem{espo96} J. A. Esposito, et al., ApJ, 461 (1996) 820. 
\bibitem{kaar96} P. Kaaret \& J. Cottam, ApJ, 462 (1998) L65. 
\bibitem{rome99} G. E. Romero, et al., A\&A, 348 (1999) 868. 
\bibitem{mukh97a} R. Mukherjee, I. A. Grenier, D. J. Thompson, AIP
                 Conf. Proc. 410 (1997) 384. 
\bibitem{gehr00} N. Gehrels, et al., Nature, 404 (2000) 363. 
\bibitem{gren00} I. A. Grenier, Nature, 404 (2000) 344. 
\bibitem{hart97} R. C. Hartman, et al., AIP Conf. Proc. 410 (1997) 307. 
\bibitem{mukh99} R. Mukherjee, et al., ApJ, 527 (1999) 132. 
\bibitem{mukh97b} R. Mukherjee, et al., ApJ, 490 (1997) 116. 
\bibitem{schl96} R. Schlickeiser, Sp. Sci. Rev., 75 (1996) 299. 
\bibitem{muec00} A. Muecke, astro-ph 0004052
\bibitem{thom00} D. J. Thompson, $\gamma$ 2000 Proc., Heidelberg. 
\bibitem{waxm00} E. Waxman, Neutrino 2000 Proc., Sudbury. 
\bibitem{hunt97} S. D. Hunter, et al., ApJ, 481 (1997) 205. 
\bibitem{sree98} P. Sreekumar, et al., ApJ, 494 (1998) 523. 
\bibitem{loeb00} A. Loeb \& E. Waxman, Nature, 405 (2000) 156. 
\bibitem{chia98} J. Chiang \& R. Mukherjee, ApJ, 496 (1998) 752. 
\bibitem{stec96} F. W. Stecker \& M. H. Salamon, ApJ, 464 (1996) 600.
\bibitem{week89} T. C. Weekes, et al., ApJ, 342 (1989) 379. 
\bibitem{week99} T. C. Weekes, et al., astro-ph/9910394. 
\bibitem{punc92} M. Punch, et al., Nature, 358 (1992) 477. 
\bibitem{gaid96} J. Gaidos, et al., Nature, 383 (1996) 319. 
\bibitem{buck96} J. H. Buckley, et al., ApJ, 472 (1996) L9. 
\bibitem{quin96} J. Quinn, et al., ApJ, 456 (1996) L83. 
\bibitem{brad97} S. M. Bradbury, et al., A\&A, 320 (1997) L5.  
\bibitem{punc97} M. Punch, et al., 25th ICRC, 3 (1997) 253. 
\bibitem{bhat97a} C. L. Bhat, IAUC (1997) 6709. 
\bibitem{haya98} N. Hayashida, et al., ApJ, 504 (1998) L71. 
\bibitem{kata99} J. Kataoka, et al., ApJ, 514 (1999) 138. 
\bibitem{sree00} P. Sreekumar, et al., AIP Conf. Proc. 510 (1999) 318. 
\bibitem{cata97} M. Catanese, et al., ApJ, 487 (1997) L143. 
\bibitem{ahar99} F. Aharonian, et al., A\&A, 349 (1999) 11. 
\bibitem{haus98} M. G. Hauser, et al., ApJ, 508 (1998) 25. 
\bibitem{puge96} J.-L. Puget, et al., A\&A, 308 (1996) L5.
\bibitem{bill98} S. D. Biller, et al., Phys. Rev. Lett., 80 (1998) 2992.  
\bibitem{stan98} T. Stanev \& A. Franceschini, ApJ, 494 (1998), L59. 
\bibitem{jung96} G. Jungman, M. Kamionkowski, \& K. Griest, Phys. Rep., 267
(1996) 195. 
\bibitem{oser99} S. Oser, et al., 26th ICRC, 3 (1999) 464. 
\bibitem{naur99} M. de Naurois, 26th ICRC, 5 (1999) 211. 
\bibitem{tume99} T. Tumer, et al., APh, 11 (1999) 271. 
\bibitem{bhat97b} C. L. Bhat, ``Towards a Major Atmospheric Cerenkov 
Detector-V,'' Kruger Park, South Africa (1997).

\bibitem{sinn95} G. Sinnis, et al., Nucl. Phys. B (Proc. Suppl.), 
                 43 (1995) 141.  
\bibitem{atki00} R. Atkins, et al., ApJ, 533 (2000) L119. 
\bibitem{atki99} R. Atkins, et al., ApJ, 525 (1999) L25. 
\bibitem{gehr99} N. Gehrels \& P. Michelson, APh, 526 (1999) 297. 
\bibitem{tava99} M. Tavani, et al., A\&AS, 138 (1999) 569. 

\end{thebibliography}
\end{document}